# Astro2020 APT White Paper

# "Mind the gap": a call to redesign astronomy graduate education

**Principal Author:**
Name: Amaya Moro-Martín
Institution: Space Telescope Science Institute
Email: amaya@stsci.edu
Phone: 410 338 2488

**Abstract:**
About one fifth of Ph.D's across all STEM disciplines secure a tenure track position in academia. It is also the case that science and engineering have evolved significantly and so has the nature of the labor market and the increasingly multidisciplinary nature of the greatest scientific challenges. These realities, however, have not altered the main objective of graduate STEM education: the training of unidisciplinary academic researchers. There is therefore a gap between what the students and society need and what graduate STEM education offers. At root of the problem is not only the lack of information regarding actual career opportunities but the lack of formation because, as the National Academies of Sciences, Engineering and Medicine recognized in its recent report entitled "Graduate STEM Education for the 21st Century", "many graduate programs do not adequately prepare students to translate their knowledge into impact in multiple careers". In that report, the National Academies of Sciences sets new standards for graduate STEM education, describes the actions required by each stakeholder, and urges all to strongly commit to this paradigm change. Astro2020 represents an opportunity for the astronomy and astrophysics community to show this commitment by issuing recommendations on the redesign of astronomy graduate education following the new guidelines set by the National Academies of Sciences. By "minding the gap" between expectations and opportunities, keeping in mind the evolving needs of the STEM workforce, we can better justify the use of tax-payers money with an investment that allows to do transformative science while conscientiously training highly qualified STEM professionals able to apply the power of science to problems and opportunities of generations to come, as envisioned by Vannevar Bush's science as the endless frontier.

# 1. Introduction

In the US, the fraction of Ph.D's across all STEM disciplines that had secured a tenure track position in academia within 5 years of graduating dropped from 27.0 % in 1993 and 25.9% in 2008 to 17.7% in 2015 (National Science Board, 2018). The situation is worse in Europe where in the United Kingdom, for example, only 3.5% of PhDs in science obtain a tenured position (Royal Society, 2010, 2014). However, most of graduate students and postdocs embark and remain in an academic career for many years with the expectation of achieving that longed-for tenured position. There is therefore a gap between the professional expectations promoted by the academic system itself and the reality of the labor market. This gap was recognized by Astro2010 and resulted in this recommendation: *"The American Astronomical Society and the American Physical Society, alongside the nation's astronomy and astrophysics departments, should make both undergraduate and graduate students aware of the wide variety of rewarding career opportunities enabled by their education, and be supportive of students' career decisions that go beyond academia. These groups should work with the federal agencies to gather and disseminate demographic data on astronomers in the workforce to inform students' career decisions."*. Almost a decade after Astro2010 the information remains sparse. The goal of this Astro2020 APT white paper is to encourage Astro2020 and the community to reflect on the idea that at root of the problem is not only the lack of information but the lack of formation because, as the National Academies of Sciences, Engineering and Medicine recognized in its recent report entitled "Graduate STEM Education for the 21st Century", *"many graduate programs do not adequately prepare students to translate their knowledge into impact in multiple careers".* The Astro2020 process represents an opportunity for the astronomy and astrophysics community to brainstorm and provide feedback on this issue and for Astro2020 to include recommendations on the redesign of astronomy graduate education following the new guidelines set by the National Academies of Sciences in its report. By "minding the gap" between expectations and opportunities, keeping in mind the evolving needs of the STEM workforce, we can better justify the use of tax-payers money with an investment that allows to do transformative science while conscientiously training highly qualified STEM professionals able to apply the power of science to problems and opportunities of generations to come, as envisioned by Vannevar Bush's science as the endless frontier.

This white paper will be submitted when the call for Astro2020 APC white papers opens. The intention of posting it earlier is to encourage the astronomy and astrophysics community to read the National Academies of Sciences report "Graduate STEM Education for the 21st Century" (in particular its Chapter 6) and to consider submitting suggestions in the form of Astro2020 APC white papers on how its recommendations could be implemented in our research field. The report will also be of interest for community members planning to submit Astro2020 APC white papers describing and assessing the successes of on-going initiatives that implement some of the recommendations in the report.

## 2. The National Academies of Sciences report on "Graduate STEM Education for the 21st Century"

There is increasing recognition that maintaining the gap between expectations and opportunities in the academic career does a disservice to many young researchers because, after many years of concatenating fixed-term contracts, sometimes more than a decade, they have to face the stigma of leaving academia having lost valuable time that could have been invested in a training more suited to their eventual job options. It also does a disservice to society because is a lost opportunity to adequately train STEM professionals capable of meeting its ever-changing needs, as *"numerous reports in the literature emphasize a lack of preparation for today's workforce, both within and outside of academia, particularly regarding communication skills, the ability to work effectively in teams, business acumen, and leadership competencies"*.

This last quote belongs to the report "Graduate STEM Education for the 21st Century" (hereafter referred to as the NASEM report for short), which was released in 2018 by the National Academies of Sciences, Engineering and Medicine. In the presentation of this report at the 2019 Annual Meeting of the American Association for the Advancement of Science (AAAS), Alan Leshner, Chair of the report (and former CEO of the AAAS and Executive publisher of the Science family of journals) explained that this study was tasked to expand on a conversation that is ongoing nationally, internationally, and across all STEM disciplines regarding the growing disconnect between what the students (and society) need and what the graduate programs offer. He argued that graduate STEM education, in essence, has not changed over a century because its main objective remains the same: the training of unidisciplinary academic researchers. However, science and engineering have evolved significantly and so has the nature of the labor market and the increasingly multidisciplinary nature of the greatest scientific challenges. Even though there is increasing recognition of this disconnect, it has been maintained because the status quo benefits most of the actors involved: the advisors and principal investigators benefit because the young researchers contribute to their scientific productivity and ability to win more funding; the academic institutions benefit because their status and funding depend partly on their scientific productivity; and the funding agencies benefit because greater scientific productivity justifies the investment made. The stakeholders that do not benefit from this status quo are the students.

The NASEM report recognizes this and argues that against the change in graduate STEM education is the incentive system that dominates in the academic culture. This includes salary promotion and tenure policies, that disproportionately favor research productivity and grants received over the quality of teaching, advising and mentoring, and grant award policies and funding criteria, that disproportionately favor research productivity even in the case of grants that involve student training. As a consequence of this incentive imbalance, academia has turned into a research-producing rather than an education-focused enterprise, resulting in lower quality teaching, advising, and mentoring. The NASEM report confronts the need for culture change by advocating for an increased emphasis on high quality teaching, advising and mentoring, for the reduction of the stigma on non-academic careers, and for the realignment of the incentive system both in academia and in federal and state funding programs.

The NASEM report recommends a systems approach with action steps for each stakeholder, emphasizing that the systemic change required to achieve the ideal graduate education it proposes (summarized below) will not be realized unless there is a sustained and robust commitment from all of them. These stakeholders include: institutions of higher education; graduate schools, departments and programs; faculty members; state and federal government agencies; private foundations and other nongovernmental organizations; employers in industry, government, and other organizations; professional societies; and graduate students. Summarized below are the key recommendations for some of these stakeholders, but readers are encouraged to read the complete summary in the Chapter 6 of the NASEM report.

**2.1 The ideal graduate education**

The NASEM report identifies the main aspects of an ideal graduate education, including:

- Ability of prospective students to select graduate programs based on easy access to outcome data on viable career pathways and successes of alumni.
- Inclusive, equitable learning and working environments that allow a diverse student population to thrive.
- Acquisition of core set of competencies. For master's programs, these include: the ability to do research; the acquisition of disciplinary and interdisciplinary knowledge and professional competencies; and the acquisition of foundational and transferrable skills such as communication, leadership and the ability to work as a team. For Ph.D. programs, these include: the realization of original research; specialized expertise in at least one STEM discipline; the ability to work in collaborative and multidisciplinary team environments; the appreciation for the ethics of the scientific enterprise; management, leadership, financial, and entrepreneurial skills; the ability to communicate to diverse audiences; and mentorship skills.
- Exposition to state-of-the-art science.
- Opportunities to understand the relationships between science, engineering, and society, to consider the broader needs of society, and the ethical and cultural issues surrounding their work.
- Opportunities to learn to communicate to diverse audiences, including other STEM professionals, policy makers and the public.
- Project-based learning opportunities, with learn-by-doing rather than lecturing and coursework as the norm.
- Opportunities to explore different career paths not only through courses and seminars, but through internships and other kinds of real-life experiences, with faculty members encouraging career exploration destigmatizing career paths outside academia. (A survey prepared by Nature in 2017 revealed that more than a third of the 5,700 doctoral students surveyed had not had any useful conversation with their thesis supervisor about career prospects and also a third said they strongly disagreed with the statement that their thesis supervisor had provided useful information on job opportunities outside academia Woolstone, 2017).
- High quality advising and mentoring from faculty adequately trained to navigate relationships in which there might be differences regarding culture and career aspirations, and to set up relationships where the goals are set jointly by the mentor and the mentee and are reviewed regularly.

- Opportunities to create an advising and mentoring network that better addresses the wide variety of needs the student will have. (This also avoids the creation of a single vertical dependency relationship that can leave the student in a vulnerable situation. This is a common recommendation to improve the climate in the academic environment and put an end to micro- and macro-aggressions to underrepresented groups (NASEM 2018b)).
- Opportunities to provide feedback to faculty and administration regarding issues important to students.

**2.2 Key recommendations for the redesign of graduate STEM education**

As stated above, the NASEM report recommends a systems approach and includes action steps for all stakeholders. For the institutions of higher education, its key recommendations include:
- The redesign of the graduate programs so that students have the opportunity to achieve the set of core competencies described in the report before they graduate.
- The improvement of the quality of teaching and mentoring, which requires the recognition of the quality of these activities in the promotion and tenure criteria, and to provide faculty members the time and the resources to learn evidence-based and inclusive teaching and mentoring practices.
- The collection and sharing of the outcome data of graduate programs including, but not limited to, completion rates, time to degree, and career outcomes and paths of alumni (the latter collected at regular time intervals for a period spanning 15 years after graduation and in a way that follows the standard definitions that correspond with national STEM education and workforce surveys). When possible, the data should be disaggregated by demographics, including gender, race and ethnicity, and visa status.
- The development and periodic evaluation of evidence-based strategies to accelerate increasing diversity and improving equity and inclusion.
- To provide information and opportunities for students to explore different career options inside and outside academia, with the relevant courses, activities, internships and other professional experiences to be developed in collaboration with industry, non-profit organizations, and other employers, where all these professional development opportunities are included in the curricula.
- To provide students with opportunities to work in teams.
- To review and periodically modify the curricula so that the graduate education programs, including their requirements for dissertations and capstone projects, reflect cutting-edge science and its collaborative and multidisciplinary nature.
- To administer periodic cultural climate surveys that assess the well-being of the graduate student population and to provide resources to contribute to their mental health. (According to a survey conducted by Nature in 2017, out of 5,700 doctoral students internationally, 45% have sought help for anxiety or depression caused by their doctoral program and 25% mention mental health among their main concerns - Nature, 550, 429, 2017).
- To develop comprehensive strategies for recruiting and retaining faculty from historically underrepresented groups in academia.

- To periodically monitor and adjust all the strategies implemented to achieve the desired objective.

The report makes additional recommendations for faculty members and for the graduate schools, departments and programs. Regarding the latter, they include: to facilitate the creation of advising and mentoring networks for students inside and outside the department (providing mentoring for the latter); to eliminate program requirements (including on dissertation and capstone projects) and curricula elements that do not provide the core competencies and learning objectives; to provide opportunities for team work and multidisciplinary learning; to collect the outcome data of the graduate program (described above); to carryout and share publicly evaluations and assessments of how the graduate program offers the core competencies; to collaborate with other STEM stakeholders outside academia on how to redesign the graduate program to reflect state-of-the art STEM activities; to adopt and periodically monitor strategies that improve on aspects of diversity, equity, and inclusion and of graduate students mental health.

The report also makes key recommendations to be implemented by the federal and state funding agencies, including: to require institutions that receive support for graduate education to collect and share the outcome data on graduate programs (described above); to realign grant award policies and funding criteria of programs that support graduate students to emphasize the quality of teaching and mentoring and to ensure the students receive the type of graduate education described in the report; to promote diversity, equity and inclusion by embedding the relevant metrics in the award criteria and by establishing accountability mechanisms; to issue calls for proposals to better understand the graduate education system and outcome of various interventions and policies.

Readers are strongly encouraged to read Chapter 6 of the NASEM report for more details on the recommendations summarized above and for the recommendations to the rest of the stakeholder (including private foundations and other nongovernmental organizations, employers in industry, government, and other organizations, professional societies, and graduate students).

## 3. On-going initiatives

Several institutions have started to address some of the recommendations summarized above, including the Council of Graduate Schools that is reviewing the basic competencies that should be provided by graduate programs; the Association of American Universities, that is advocating for the collection and publication of career outcome data; and the National Institutes of Health, that has released a [funding opportunity][1] to promote the development and improvement of graduate programs in the life sciences, following many of the standards proposed in the report. In Europe, the European Research Council has started collecting data on what happens with the scientific careers of the researchers who have benefited from its grants, including the researchers hired under its funded programs. Rolf Tarrach, the President of the European Association of Universities, is also in favor of European universities compiling career outcome data and making them public but doubts this will happen unless there is pressure from national and European institutions to do it.

---

[1] Predoctoral Institutional Research Training Grant (T32): https://grants.nih.gov/grants/guide/pa-files/PAR-17-341.html

The life sciences are a research area where the gap between expectations and opportunities is more pronounced, with only 10% of Ph.D's in the US able to secure a tenure track position within 5 years of graduating. This might be the case why it is in these disciplines where more concrete initiatives are crystalizing, including the funding opportunity by the National Institutes of Health mentioned above, and the formation of a coalition of universities committed to collecting and sharing career outcome data[2] on their graduate programs in the life sciences. The goals of this latter effort, as described by Blank et al. (2017), are: to allow prospective students to make better informed decisions in terms of the choice of graduate programs; to allow graduate students and postdoctoral researchers to make decisions regarding their careers long before they are doing it now; to serve as a reality check for the institutions to assess the effectiveness of their training programs, allowing their redesign based on evidence; and to enable accountability mechanism within the institutions themselves and with regard to funding agencies regarding their success to prepare students to actual career outcomes.

Environmental science is a research area where communication training is important because there is the need to redefine and execute a scientific agenda able to address urgent and unprecedented environmental and social challenges and science communication plays a key role in enabling this. Jane Lubchenco, former Administrator of NOAA, in her 1997 address as President of the AAAS, focused on the need of a new social contract for science by which scientist are committed not only create new knowledge that is helpful to society but to share it widely, addressing the growing disconnect between science, policy makers and the public. And to enable this she cofounded three organizations to help scientists become better communicators. One of them is COMPASS[3], a non-profit organization that since 1999 has been helping environmental scientists to enter the conversations that result in the decision making, to make sure their scientific work has a real impact on society and does not simply sit on journals. All STEM professionals should have access to the body of knowledge on how to communicate effectively to diverse audiences, including the barriers policy makers face when evaluating scientific data (e.g. lack of time, difficulties in accessing information, difficulties in interpreting the results, differences between the time scales in politics and science, lack of relevance, lack of credibility, etc.) and the ways scientific evidence is typically used. Non-profit organizations like COMPASS could aid academic institutions to introduce this aspect of science communication into their graduate education curricula, ensuring all STEM professionals have a strong foundation. This is important not only in the environmental sciences but in all STEM disciplines, including Astronomy and Astrophysics, as the quest of new knowledge needs to be part of the agenda.

## 4. Concluding remarks

Cultural shifts are hard to realize and, as the NASEM report emphasizes, the change of paradigm regarding graduate STEM education will not happen unless all stakeholders show a firm, common, and commensurate commitment. Astro2020 represents a unique opportunity for the astronomy and astrophysics community to show this commitment by making recommendations on how to redesign astronomy graduate education following the NASEM report's guidelines. By "minding the gap" between expectations and opportunities, keeping in mind the evolving needs of the STEM workforce, we can better justify the use of tax-payers money with an investment that allows to do transformative science while conscientiously training highly qualified STEM professionals able to apply the power of science to problems and opportunities of generations to come, as envisioned by Vannevar Bush's science as the endless frontier.

---

[2] This is an example of what type of data is being collected and shared: https://graduate.ucsf.edu/graduate-program-statistics
[3] COMPAS: https://www.compassscicomm.org